\documentclass[intlimits,twoside,a4paper]{article}

\usepackage{amsmath,amssymb}
\usepackage{graphicx}

\usepackage[T2A]{fontenc}
\usepackage[cp1251]{inputenc}

\def\be{\begin{equation}}
\def\ee{\end{equation}}

\usepackage[eqsecnum]{cmpj2}
\issue{2013}{16}{4}{43605}
\doinumber{10.5488/CMP.16.43605}

\title[Two-dimensional core-softened model]%
{Two-dimensional core-softened model with water like properties. Study by thermodynamic perturbation theory%
\thanks{Dedicated to Professor Myroslav Holovko on the occasion of his 70$^\textrm{th}$ birthday.}}
\author[T. Urbic]{T. Urbic}
\address{Faculty of Chemistry and Chemical Technology,
University of Ljubljana, Askerceva 5, 1000 Ljubljana, Slovenia}

\date{Received August 1, 2013, in final form August 28, 2013}
\authorcopyright{T. Urbic, 2013}

\begin{document}

\maketitle

\begin{abstract}
Thermodynamic properties of the particles interacting through smooth version of Stell-Hemmer interaction were studied using Wertheim's thermodynamic perturbation theory. The temperature dependence of molar volume, heat capacity, isothermal compressibility and thermal expansion coefficient at constant pressure for different number of bonding sites on particle were evaluated. The model showed water-like anomalies for all evaluated quantities, but thermodynamic perturbation theory does not properly predict the dependence of these properties at a fixed number of bonding points.
\keywords Monte Carlo, thermodynamic perturbation theory, core softened fluid
\pacs 64.70.Ja, 61.20.Ja
\end{abstract}

\section{Introduction}

They were Stell and Hemmer who first proposed core-softened potentials in 1970~\cite{stell1}. In their early work, they stressed that negative curvature in interaction potential might lead to a second critical point in addition to a standard liquid-gas critical point. In different works~\cite{barbosa,sharma} it has been shown that core-softened potentials and similar shouldered potentials can reproduce various fluid anomalies that are typical of water and other substances with angular dependent interactions, such as silica~\cite{sharma}, silicon~\cite{sastry}, BeF$_2$~\cite{angell}. Core-softened potentials were also used to study single-component liquid metal systems~\cite{l1,l2,l3,l4,l5} and as solvent for studying ions~\cite{miha}. Poole et al.~\cite{poole} proposed liquid-liquid phase transition as an explanation for anomalous properties of water.
After that there was an increased interest to the studies of these liquid-liquid phase transitions. Franseze et al.~\cite{franz} suggested that the liquid-liquid phase transition and its critical point might be caused by the potential with two characteristic distances (hard core and soft core). In their work, they reported the existence of the low-density liquid phase and the high-density liquid phase obtained for 3D model using molecular dynamics (MD) simulations. On the other hand, 2D MD produced only a density anomaly but no liquid-liquid phase transition~\cite{rsl1,rsl2}. Scala et al.~\cite{scala1} carried out MD simulations of 2D discrete and smoothed version of potential to study liquid anomalies. These studies were continued by Buldyrev et al.~\cite{buly1} to explore liquid-liquid phase transition for 2D and 3D version of potentials and by Almudallal et al.~\cite{al}. They both produced phase diagrams for a discrete version of potential with liquid anomalies, and no liquid-liquid critical point in stable liquid region was obtained.

Our aim here is to apply Wertheim's thermodynamic perturbation theory (TPT) to capture the physics of the model of 2D molecules interacting by Stell-Hemmer potential.
In recent years, a theory has been developed for fluids comprised of molecules that associate into dimers and higher clusters due to the presence of highly directional attractive forces~\cite{w34,w34_,wtpt}. In the present work, we apply the thermodynamic perturbation theory (TPT)~\cite{w34,w34_,wtpt,hol} to central symmetric attractive potential.

\section{Model}

\begin{figure}[htb]
\centerline{
\includegraphics[height=0.5\textwidth, angle=270]{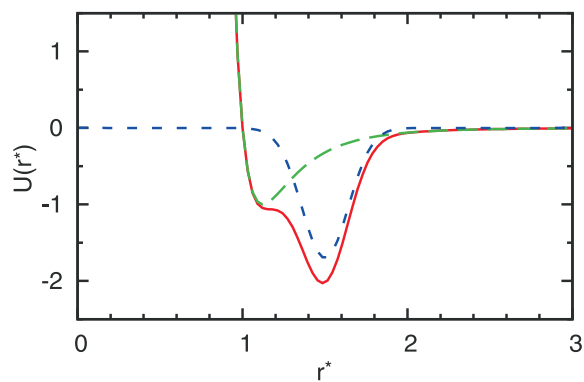}
}
\caption{(Color online) The core-softened potential $U(r)$ (solid line) with both contributions (LJ~--- long dashed line and Gaussian part~--- dashed line).}
\label{fig:sh.pot}
\end{figure}

The smooth version of the core-softened potential proposed by Scala et al.~\cite{scala1} is used in this work. The interaction potential $U(r)$ is a sum of a Gaussian well and the Lennard-Jones (LJ) part of the potential
\be
U(r)=U_\textrm{LJ}(r)+U_a(r),
\label{int1}
\ee
where $U_\textrm{LJ}(r)$ is standard Lennard-Jones potential
\be U_\textrm{LJ}(r)=4\varepsilon\left[\left({\frac{\sigma}{r}}\right)^{12}
-\left({\frac{\sigma}{r}}\right)^6\right]. \label{lj1}
\ee
$\varepsilon$ is here the well-depth and $\sigma$ is the distance, where LJ part of the potential is zero. The Gaussian part of the interaction is as follows:
\be
U_a(r)=-\lambda\varepsilon\exp{\left[-a\left( \frac{r-r_0}{\sigma}\right)^2\right]}.
\label{int2}
\ee
This part of potential is stronger than the LJ part and is the reason that the particles make strong association. We also refer to this part of potential as association potential. We apply the units and values of model parameters as used before by Scala et al.~\cite{scala1} as $\varepsilon=1.0$, $\sigma=1.0$, $\lambda=1.7$, $a=25.0$, $r_0= 1.5\sigma$. Figure~\ref{fig:sh.pot} shows the shape of the smooth version of the core-softened potential used here.

\section{Monte Carlo simulation details}

We performed Monte Carlo simulations in the
isothermal-isobaric (NpT) ensemble to obtain thermodynamic properties of the model.
At each step, the displacements in the $x,y$ coordinates were chosen randomly. We used periodic boundary conditions and the minimum image convention to mimic an infinite system of particles. The starting configurations were selected at random. Every 10 moves of particles an attempt is made to scale the dimensions of the box and all of its component particles in order to hold the pressure constant. 5$\times$10$^4$ moves per particle were needed to equilibrate the system.  The statistics were gathered over the next 1$\times$10$^6$ moves to obtain well converged results. All simulations were performed with $N=$200 or $N=$400 molecules. The maximum change of dimensions of the box was calibrated during equilibration simulations. The physical properties of the system such as enthalpy and volume were calculated as the statistical averages of these quantities over the course of  simulations~\cite{frenkel}. The heat capacity, $C_p$, the isothermal compressibility, $\kappa$, and the thermal expansion coefficient, $\alpha$ are computed from the fluctuations~\cite{mbs} of enthalpy, $H$, and volume, $V$.

\begin{align}
C_{p}&=\frac{C_p}{k_\textrm{B}}=\frac{\langle H^2\rangle -\langle H\rangle ^2}{NT^2}\, ,\nonumber\\
\kappa&= \frac{\langle V^2\rangle -\langle V\rangle ^2}{T\langle V \rangle }\, ,\nonumber\\
\alpha&=\frac{\langle VH\rangle-\langle V\rangle \langle H \rangle }{T^2\langle V\rangle }\, ,
\label{eq:fluctuations}
\end{align}
$T$ is temperature of the system and $N$ number of particles.

\section{Thermodynamic perturbation theory}

The Helmholtz free energy of the system is the key quantity of the thermodynamic perturbation theory~\cite{w34,w34_}. In case of the model studied in this work, this
quantity is the sum of two terms
\be {\frac{A}{Nk_\textrm{B}T}}={\frac{A_\textrm{LJ}}
{Nk_\textrm{B}T}}+{\frac{A_{a}}{Nk_\textrm{B}T}}\,. \label{tptfree}
\ee
$N$ is the number of molecules, $T$ is temperature and $k_\textrm{B}$
is Boltzmann's constant.  The Helmholtz free energy of
Lennard-Jones system, $A_\textrm{LJ}$, is calculated using the Barker-Henderson perturbation theory~\cite{hand1}
\be {\frac{A_\textrm{LJ}}{Nk_\textrm{B}T}}={\frac{A_\textrm{HD}}
{Nk_\textrm{B}T}}+{{\frac{\rho}{2k_\textrm{B}T}} \int_{\sigma}^ \infty
g_\textrm{HD}(r,\eta)U_\textrm{LJ}(r)\rd {r}}. \label{alj1}
\ee
${A_\textrm{HD}}$ is the hard-disk contribution to the Helmholtz free
energy, $g_\textrm{HD}(r,\eta)$ is pair correlation function for hard disks at packing fraction $\eta={\frac{1}{4}} \pi d^2\rho$ and $\rho$ is the number density of molecules. $d$ is the hard-disk diameter calculated using Barker-Henderson approximation as
\begin{equation}
d = \int_0^{\sigma}  \left [1 - \exp \left(-\frac{U_\textrm{LJ}}{k_\textrm{B}T}\right ) \right ]\rd r.
\label{eq:dHS}
\end{equation}
We used the procedure by Scalise et al.~\cite{scalise} to calculate the HD term of the Helmholtz free energy
\be
{\frac{A_\textrm{HD}-A_\textrm{ideal}}{Nk_\textrm{B}T}}=-1.10865-0.8678\ln{(1-
\eta)}-0.0157(1-\eta)+{\frac{1.1322}{1-\eta}}-{\frac{0.00785}{({1-
\eta})^2}}\,. \label{ahd1}
\ee
For $g_\textrm{HD}(r)$, the expression of Gonzalez et al.~\cite{gonzalez} was used.

The association contribution to Helmholtz free energy, $A_a$, was calculated by~\cite{jackson,w34,w34_}
\be {\frac{A_{a}}{Nk_\textrm{B}T}}=N_a{\left(\log{x}-{\frac{x}{2}}+{\frac{1}{2}}\right)}, \label{ahb2}
\ee
where $x$ is the fraction of molecules not bonded at particular interaction site and is obtained from the mass-action law~\cite{w34,w34_} in the form
\be {x}={\frac{1}
{1+{N_a\rho x\Delta}}}\,. \label{mass1}
\ee
$\rho$ is the
total number density. Finally, $\Delta$ is defined by
~\cite{w34,w34_,jackson}
\be {\Delta}={2\pi\int{g_\textrm{LJ}(r)
f_a(r)r\rd r}}. \label{povp}
\ee
$f_a(r)$ is a Mayer function for the association
potential
\be
f_a(r)=\exp{\left[-\frac{U_a(r)}{k_\textrm{B}T}\right]}-1.
\ee
The pair distribution function $g_\textrm{LJ}(r)$
is obtained by solving the Percus-Yevick equation for
Lennard-Jones disks. $N_a$ is the number of association points on the particles. Particles have a spherically symmetric association potential. They do not have a varied number of bonding points as is usually the case where Wertheim's theory is used. We made approximation that each particle can have $N_a$ association points in the center of a particle interacting with associating potential. We used a different number of interaction sites, from 1 to 6, the last being a coordination number of particle in a perfect hexagonal crystal. Once the Helmholtz free energy is known, other thermodynamic quantities may be calculated from standard thermodynamic relations~\cite{hand1}
\be
p=\frac{\rho^2}{N}\left(\frac{\partial A}{\partial \rho}\right)_T \,,
\ee
\begin{eqnarray}
(\kappa_T)^{-1} = \rho \left(\frac{\partial P}{\partial \rho}\right )_{T}\, ,
\label{eq:other_thermo_relations1}
\end{eqnarray}
\begin{eqnarray}
\alpha = \kappa_T \left(\frac{\partial P}{\partial
    T}\right )_{\rho}\, ,
\label{eq:other_thermo_relations2}
\end{eqnarray}
\begin{eqnarray}
C_P = C_V + \frac{\alpha}{\rho}\left [P - \rho^2 \left (\frac{\partial
      U}{\partial \rho} \right )_{T} \right ]\,.
\label{eq:other_thermo_relations3}
\end{eqnarray}

\section{Results}

%\begin{figure}[htb]
%\centerline{
%\includegraphics[width=80mm]{alfa}
%}
%  \caption{(Color online) Temperature dependence of the thermal expansion
%coefficient at $P^* = 0.75 $; legend as for figure~\ref{fig:molv}.}
%  \label{fig:alfa}
%\end{figure}

All the results are given in reduced units; the excess internal energy and temperature are normalized to the LJ interaction parameter $\varepsilon$ ($E^*={{E}/{\varepsilon}}$, $T^*={{k_\textrm{B}T}/{\varepsilon}}$) and all the distances are scaled to the characteristic length $\sigma$ ($r^*={{r}/{\sigma}}$).

\begin{figure}[htb]
\centerline{
\includegraphics[width=0.5\textwidth]{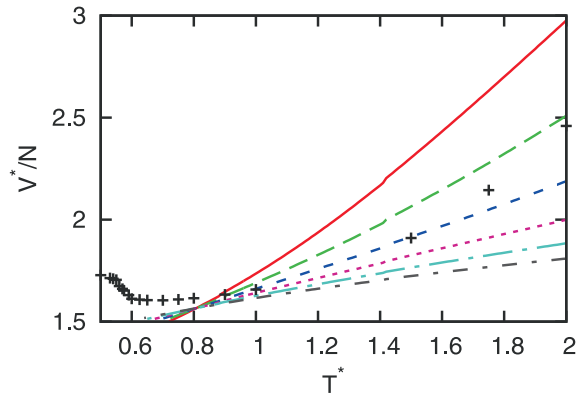}
}
  \caption{(Color online) Temperature dependence of the molar volume at $P^* =
0.75 $ as obtained by the Monte Carlo simulation (symbols), the
thermodynamic perturbation theory for a different number of associating points ($N_a=1$ full red line, $N_a=2$ long dashed green line, $N_a=3$ dashed blue line, $N_a=4$ dotted pink line, $N_a=5$ long dashed-dotted light blue line and $N_a=6$ dashed-dotted grey line.}
  \label{fig:molv}
\end{figure}
In figure~\ref{fig:molv}, we compare the molar volume or volume per particle, $V^*/N$, obtained from the Monte Carlo simulations, with the results of the thermodynamic perturbation theory for a different number of bonding sites on particle ($N_a=1 - 6$). The calculations were performed at a reduced pressure of $P^* = 0.75$. We find out that the TPT does not properly capture the results for simulations as well as it does not predict the maxima in density or minima in molar volume. In order to obtain an agreement, $N_a$ should be dynamically varied. $N_a$ should be varied with temperature and density in order to get an agreement between theoretical and simulation results. We can see from figures that we have good agreement of TPT with simulation for $N_a=2$ at high temperature $T^*=2.0$, then 3 at $T^*=1$, etc.

\begin{figure}[htb]
\centerline{
\includegraphics[width=0.48\textwidth]{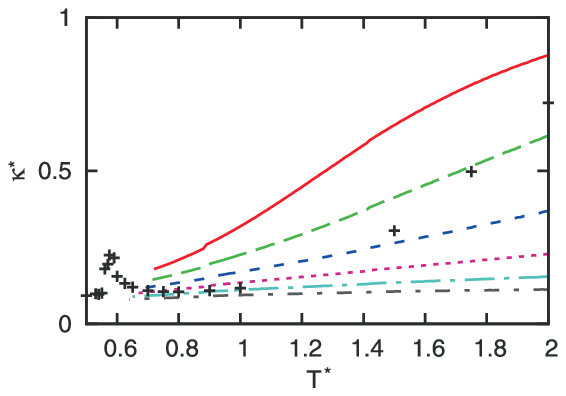}
\hspace{2mm}%
\includegraphics[width=0.48\textwidth]{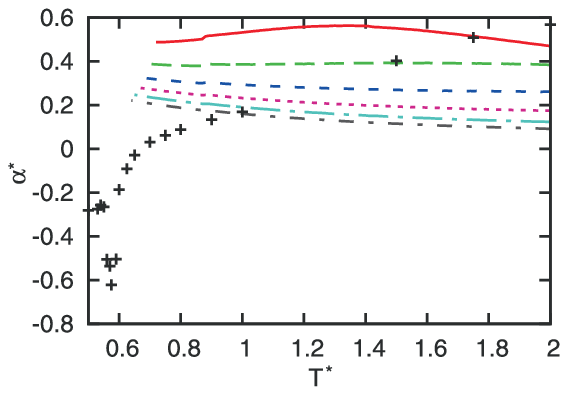}
}
\parbox[t]{0.5\textwidth}{
\caption{(Color online) Temperature dependence of the isothermal compressibility at $P^* = 0.75 $; legend as for figure~\ref{fig:molv}.\label{fig:kapa}}
}
\parbox[t]{0.5\textwidth}{
  \caption{(Color online) Temperature dependence of the thermal expansion
coefficient at $P^* = 0.75 $; legend as for figure~\ref{fig:molv}.  \label{fig:alfa}}
}
\end{figure}

\begin{figure}[htb]
\centerline{
\includegraphics[width=0.48\textwidth]{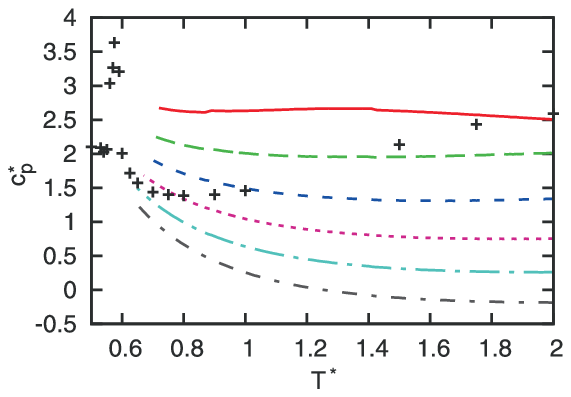}
\hspace{2mm}
\includegraphics[width=0.48\textwidth]{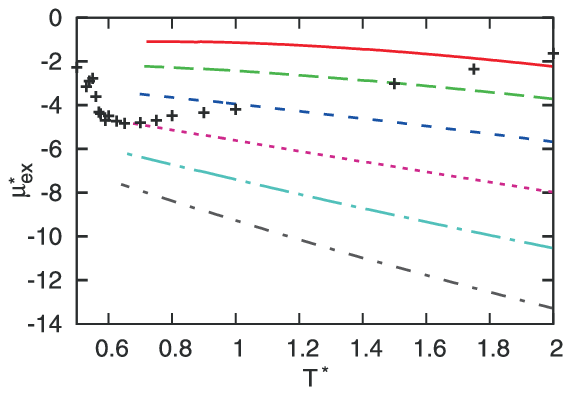}
}
\parbox[t]{0.5\textwidth}{
\caption{(Color online) Temperature dependence of the heat capacity at $P^* = 0.75 $; legend as for figure~\ref{fig:molv}.}
  \label{fig:cp}
}
\parbox[t]{0.5\textwidth}{
  \caption{(Color online) Temperature dependence of the excess chemical potential at $P^* = 0.75 $; legend as for figure~\ref{fig:molv}.}
  \label{fig:mu}
}
\end{figure}

The remaining figures show the temperature dependencies of the other
thermodynamic quantities of interest: the isothermal
compressibility, $\kappa_T^*$ (figure~\ref{fig:kapa}), the thermal expansion
coefficient, $\alpha^*$ (figure~\ref{fig:alfa}), the heat capacity, $C_p^*$
(figure~\ref{fig:cp}), and the excess chemical potential $\mu^*_\textrm{ex}$ (figure~\ref{fig:mu}).
The TPT for a fixed number of associating points
is not in agreement with the Monte Carlo simulation data for all quantities. From the results we can also see that the model behaves in a way that the number of associating points is not fixed, but it changes at a constant pressure with temperature. We calculated free energy as a function of the number of boding sites, $N_a$, and the free energy  decreases within the whole range.

%\begin{figure}[htb]
%\centerline{
%\includegraphics[width=80mm]{mu}
%}
%  \caption{(Color online) Temperature dependence of the excess chemical potential at $P^* = 0.75 $; legend as for figure~\ref{fig:molv}.}
%  \label{fig:mu}
%\end{figure}

\section{Conclusion}

The thermodynamic perturbation theory was used to
study the thermodynamics of the particles interacting through a smooth version of Stell-Hemmer interaction. The results for the molar volume, the isothermal compressibility, the thermal expansion coefficient, the heat capacity and the excess chemical potential obtained by the TPT theory for a fixed number of bonding sites on the particles are not in agreement with the computer simulation
results for all the parameters studied. This is caused by the fact that the interaction is spherical symmetric, and for TPT we have to make approximations. It is crucial to change the spherical symmetric potential into a directional one and to have a different number of interaction points with directional forces. We cannot obtain correct thermodynamic properties with a fixed number of association points. Proper thermodynamics could be obtained if the number of association points  varied with temperature and pressure.

\section{Acknowledgements}

We appreciate the support by the Slovenian Research Agency
(P1 0103-0201 and J1 4148) and NIH Grant GM063592.

\newpage

\ukrainianpart

\title{Двовимірна модель потенціалу з м'яким кором з властивостями подібними до води. Дослідження методом термодинамічної теорії збурень}
\author{Т. Урбіч}
\address{Факультет хімії і хімічної технології, Університет м.
Любляна, 1000 м. Любляна, Словенія}

\makeukrtitle

\begin{abstract}
\tolerance=3000%
Термодинамічні властивості системи  частинок, що взаємодіють за
допомогою зм'якшеного потенціалу типу Стелла-Хеммера досліджено в
рамках термодинамічної теорії збурень Вертхайма. Розраховано
температурну залежність молярного об'єму, теплоємності, ізотермічної
стисливості і коефіцієнта термічного розширення при сталому тиску
для різного числа зв'язків для заданої частинки. Модель
характеризується  аномальною поведінкою властивостей, подібною до
води, але термодинамічна теорія збурень незадовільно описує ці
властивості, якщо число зв'язків в розрахунку на частинку є
зафіксоване.
\keywords Монте Карло, термодинамічна теорія збурень, плин з м'яким
кором
\end{abstract}

\lastpage

\end{document}